# Unveiling the Structural Origin of the High Carrier Mobility of a Molecular Monolayer on Boron Nitride

Rui Xu[1,†], Daowei He[2,†], Yuhan Zhang[2], Bing Wu[2], Fengyuan Liu[2], Lan Meng[1], Jun-Fang Liu[1], Qisheng Wu[3], Yi Shi[2], Jinlan Wang[3], Jia-Cai Nie[1], Xinran Wang[2,*], Lin He[1,*]

[1]Department of Physics, Beijing Normal University, Beijing, 100875, People's Republic of China

[2]National Laboratory of Solid State Microstructures, School of Electronic Science and Engineering, and Collaborative Innovation Center of Advanced Microstructures, Nanjing University, Nanjing, 210093, People's Republic of China

[3]Department of Physics, Southeast University, Nanjing, 211189, People's Republic of China

**Very recently, it was demonstrated that the carrier mobility of a molecular monolayer dioctylbenzothienobenzothiophene ($C_8$-BTBT) on boron nitride can reach 10 cm$^2$/Vs, the highest among the previously reported monolayer molecular field-effect transistors. Here we show that the high-quality single crystal of the $C_8$-BTBT monolayer may be the key origin of the record-high carrier mobility. We discover that the $C_8$-BTBT molecules prefer layer-by-layer growth on both hexagonal boron nitride and graphene. The flatness of these substrates substantially decreases the $C_8$-BTBT nucleation density and enables repeatable growth of large-area single crystal of the $C_8$-BTBT monolayer. Our experimental result indicates that only out-of-plane roughness greater than 0.6 nm of the substrates could induce disturbance in the crystal growth and consequently affect the charge transport. This information would be important in guiding the growth of high-quality epitaxy molecular film.**

# I. INTRODUCTION

Organic molecular crystals have many promising applications in electronics and photonics for the characteristics of flexibility, diaphaneity, and low-cost [1-8]. However, the low charge carrier mobility blocks the use of the organic molecular crystals in electronic applications [8-18]. The carrier mobility becomes even worse with decreasing the thickness of the organic crystals. One example is that the carrier mobility of monolayer organic field-effect transistors (OFETs) is only approaching 0.1 cm$^2$/Vs so far [14], although many methods have been developed to improve it in the past few years. Until very recently, we show that the carrier mobility of dioctylbenzothienobenzothiophene ($C_8$-BTBT) monolayer epitaxially grown on boron nitride could reach as high as 10 cm$^2$/Vs [19], which is comparable to that of some two-dimensional (2D) atomic crystals, such as $MoS_2$ [20-22]. This 2D molecular crystal shows great promise for the low-cost and flexible electronics applications.

To fully understand the origin of the high carry mobility of the $C_8$-BTBT monolayer, here the structures of the 2D $C_8$-BTBT crystal were carefully studied using atomic force microscopy (AFM) and cryogenic scanning tunneling microscopy (STM). Our experimental result indicates that the $C_8$-BTBT molecules prefer to grow layer-by-layer on hexagonal boron nitride and graphene, and it is facile to control the thickness of the $C_8$-BTBT crystals by adjusting the growth time and temperature. We also demonstrate that only out-of-plane roughness greater than 0.6 nm of the substrates could act as the $C_8$-BTBT nucleation center. In our experiment, the $C_8$-BTBT nucleation density is substantially reduced because of the flatness of substrates. This enables us to grow large-area single crystal of the $C_8$-BTBT layers repeatably. The large-area single crystal nature of the $C_8$-BTBT monolayer reduces disturbances in electronic transport and may be the key reason of the observed record-high carrier mobility.

## II. EXPERIMENTAL METHODS

The $C_8$-BTBT layers were synthesized by heating $C_8$-BTBT powder to 100-120 $^o$C in a tube furnace under high vacuum. Graphene and boron nitride (BN) are ideal substrates for the growth of high quality molecular film because that both of them are atomically flat without dangling bonds. In our experiments, three different substrates, i.e., graphene on $SiO_2$ (graphene/$SiO_2$), BN on $SiO_2$ (BN/$SiO_2$), and graphene on Cu foil (graphene/Cu), are used to grow the $C_8$-BTBT layers. The graphene/$SiO_2$ and BN/$SiO_2$ are obtained via transfering an exfoliated graphene and BN onto $SiO_2$ substrate, respectively. Similar observations about the growth mechanism and the structure of the $C_8$-BTBT layers are obtained on the two different substrates. The graphene/Cu foil is obtained via a traditional chemical vapor deposition (CVD) mehod [23] (See Supporting Material [24] for details), and only the $C_8$-BTBT layers grown on this substrate are further characterized by STM for the requirement of conductivity in the STM measurements. In the growth of $C_8$-BTBT, the substrate is put a few inches away from the powder. By controlling the heating temperature and duration, high-quality $C_8$-BTBT layers with different thickness can be observed [19].

Two independent AFM: an Asylum Cypher and a Veeco Multimode 8, were used under ambient condition in this work. The same result about the thickness of $C_8$-BTBT films is obtained based on the two different AFM systems. The STM system was an ultrahigh vacuum four-probe scanning probe microscope from UNISOKU. All the STM and scanning tunneling spectroscopy (STS) measurements were performed in an ultrahigh vacuum chamber ($10^{-10}$ Torr) and all the images were taken in a constant-current scanning mode at liquid-nitrogen temperature. The STM tips were obtained by chemical etching from a wire of Pt(80%) Ir(20%) alloys. Lateral dimensions observed in the STM images were calibrated using a standard graphene lattice. The tunneling spectrum, i.e., the *dI/dV-V* curve, was carried out with a standard lock-in technique using a 789-Hz alternating current modulation of the bias voltage.

## III. EXPERIMENTAL RESULTS AND DISCUSSION

To explore the growth mechanism, we intentionally interrupt the growth of the $C_8$-BTBT layers to carry out AFM measurements. The $C_8$-BTBT films are stable against air exposure. Therefore, the frequent interruption and ambient exposure of the sample for characterization do not change the morphologies of the $C_8$-BTBT layers. Figure 1(a)-(e) show sequential AFM snapshots of the $C_8$-BTBT layers grown on graphene/$SiO_2$ during a 4-minute growth. In our experiment, the $C_8$-BTBT is observed to grow only on the graphene or BN (See Figures S1 and S2 in Supporting Material [24] for more experimental results) because of very small binding energy between $C_8$-BTBT and $SiO_2$. For simplify,

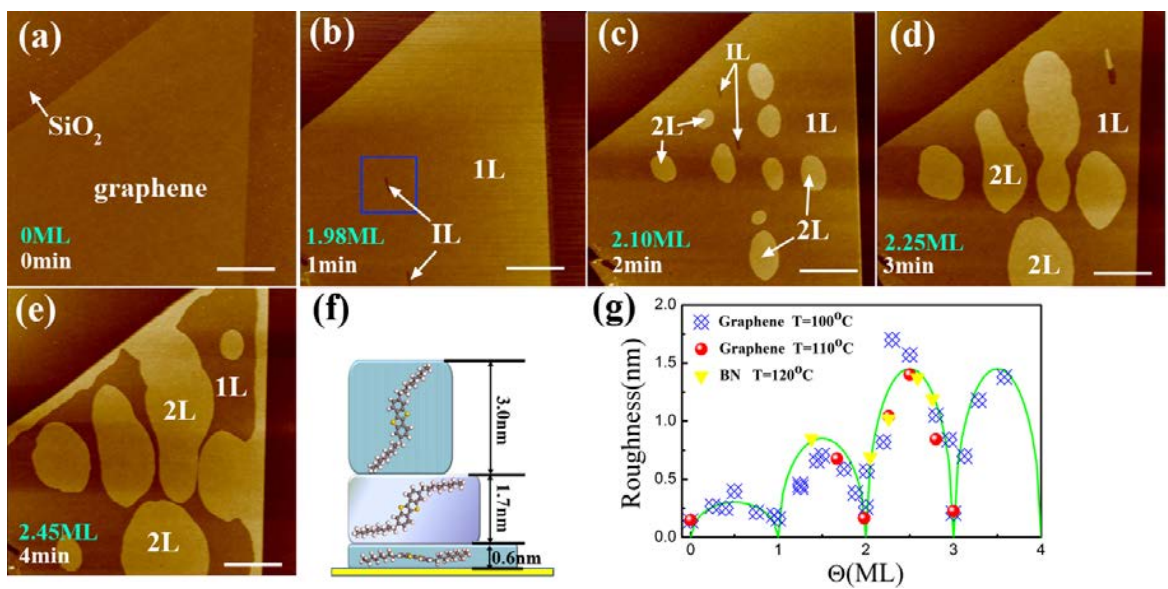

**Figure 1** (color online). (a)-(e) Sequential AFM snapshots of $C_8$-BTBT layers on graphene/$SiO_2$ during a 4-minute growth. The scale bars are 2 μm. The thickness of molecular film and the duration of growth are marked on each image. In panels (b) and (c), the position indicated by the white arrows and IL are regions only covered by the IL. (f) The schematic diagram of the molecular structure of $C_8$-BTBT layers on graphene or BN. The thicknesses of the interfacial layer (IL), first layer (1L), and second layer (2L) on both substrates are 0.60±0.08 nm, 1.70±0.09 nm, and 3.00±0.15 nm, respectively. (g) The roughness of $C_8$-BTBT layers as a function of the coverage. The green curve is calculated according to Eq. (1). The red solid circles and blue symbols are experimental data on graphene recorded at the growth temperature 110 °C and 100 °C, respectively. The yellow triangles are experimental data of $C_8$-BTBT layers on BN.

we introduce coverage $\Theta$, which represents thickness of the $C_8$-BTBT layers, to describe our experimental result. The coverage is expressed in monolayer (ML) units and can be measured from AFM images [14]. For example, if one layer $C_8$-BTBT covers 98% of the whole graphene substrate, we obtain the coverage of sample as $\Theta = 0.98$ ML, and we define $\Theta = 1.0$ ML when one layer $C_8$-BTBT covers the graphene surface completely. At growth temperature 110 $^o$C, the thickness of sample reaches $\Theta = 1.98$ ML after 1-minute growth. Then, the initial two layers can be observed on graphene, i.e., the interfacial layer (IL) covers the whole graphene region and the first layer (1L) covers 98% of the whole graphene region (Fig. 1(b) and also see Figure S3 in Supporting Material [24]). The second layer (2L) begins to nucleate before the completion of the 1L (Fig. 1(c)). After 4-minute growth, the 1L layer covers the whole graphene and the 2L layer almost covers half of it (Fig. 1(e)). Our experimental result indicates that the $C_8$-BTBT grows in a molecularly ordered layer-by-layer fashion on the graphene/$SiO_2$ [25].

Other samples grown on graphene/$SiO_2$ and BN/$SiO_2$ at different temperatures are also investigated, similar morphologies and thickness of the $C_8$-BTBT layers are obtained (See Figures S1 and S2 in Supporting Material [24] and more data in Ref. [19]), indicating similar microscopic structures and growth mechanism of the $C_8$-BTBT layers on the two substrates. Figure 1(f) shows schematic structures of the initial three $C_8$-BTBT layers grown on the graphene/$SiO_2$ or BN/$SiO_2$. The thicknesses of the IL, 1L, and 2L on both the substrates are $\Delta h_1 = 0.60 \pm 0.08$ nm, $\Delta h_2 = 1.70 \pm 0.09$ nm, and $\Delta h_3 = 3.00 \pm 0.15$ nm respectively. For Θ ≥ 3.0 ML, the thickness of each layer is about 3.0 nm consisting with the c-axis length of the primitive cell in the bulk crystal [26-29]. The thickness of 0.6 nm for the IL arises from a new form of molecular packing, as shown subsequently, attributing to the van der Waals (vdW) interaction between substrates (graphene or BN) and the $C_8$-BTBT, which allows large lattice mismatch betwen the substrate and film [30,31]. The vdW forces decay rapidly as $r^{-6}$, therefore, the role of substrate is much reduced in the 1L and becomes negligible in the 2L and above [32].

In order to quantitatively understand the growth mechanism of the $C_8$-BTBT layers, the evolution of the morphology has been further analyzed by means of the surface roughness *w*, which describes the out-of-plane disorder with respect to the homogeneous layer in a layered morphology. The *w* is the root mean square fluctuation of the film topography *h* and, in layer-by-layer growth, it can be expressed as [14]:

$$w = \sqrt{\langle h^2 \rangle - \langle h \rangle^2} = \Delta h_n [(2n-1)\Theta - n(n-1) - \Theta^2]^{1/2}. \qquad (1)$$

Here $\Delta h_n$ is the thickness of the $(n+1)th$ layer (According to our definition, $n = 1$ for IL, $n = 2$ for 1L, and so on). As shown in Fig. 1(g), the roughness *w* versus coverage Θ measured on both the graphene and BN in our experiment is in good agreement with the predictions of Eq. (1). Therefore, our experimental result demonstrates that the $C_8$-BTBT molecular crystals prefer layer-by-layer growth for the initial few layers on the two substrates. This result may also be valid for other substrates that are atomically flat without dangling bonds [31].

The structure of the $C_8$-BTBT layers is further studied by STM measurement. However, only the structure of the IL could be clearly derived from STM images because of limited vertical conductivity of the molecular layers. In our experiment, the IL plays an important role for the observed record-high carry mobility. It not only separates the $C_8$-BTBT 1L from the influence of the substrate but also becomes the new "substrate" for the growth of the 1L. With carefully control the flatness of the substrate, i.e., the graphene on Cu foil (Fig. 2(a)), we successfully synthesize large-area single crystal IL, as shown in Fig. 2(b). It is interesting to observe that the growth of the IL is not disturbed by small steps of the substrate. Higher magnification STM image of the IL is shown in Fig. 2(c). The $C_8$-BTBT molecules are packed in a rectangular lattice with a period of 2.52 nm and 0.66 nm in two orthogonal directions. Based on the observed STM images and density functional theory (DFT) calculations [19], an energy-minimized molecular configuration on graphene is shown in inset of Fig. 2(c) and Fig. 2(d). Figure 2(e) and 2(f) show the detailed comparison between the structure obtained by STM and the configuration generated by DFT calculation. Obviously, they agree with each other quite well. Importantly, the thickness of IL obtained by STM measurements and DFT calculation consists well with that ~ 0.6 nm acquired by AFM experiments. Here, we should point out that the structure of the IL can not be obtained by cutting along any crystallographic plane of the bulk phase of the $C_8$-BTBT crystal (See Figures S4 in Supporting Material [24]). Figure 2(g) shows a *dI/dV-V* curve of the $C_8$-BTBT IL. The tunneling spectrum reveals a typical semiconducting behavior, which

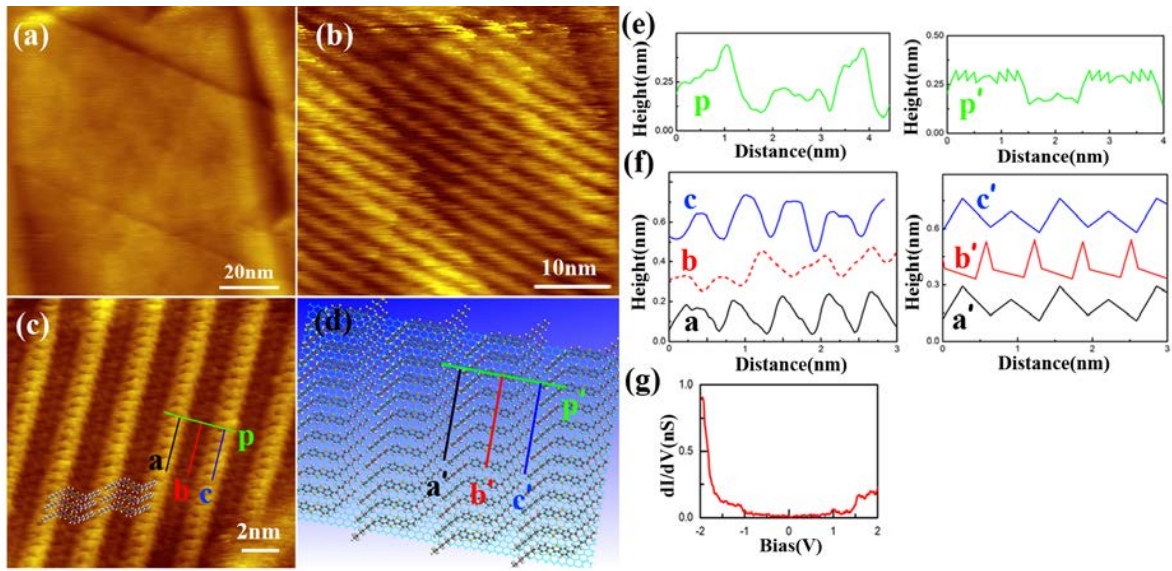

**Figure 2** (color online). (a) A large area STM image of graphene surface on Cu foil ($V_{sample}$ = -1 V and I = 16.8 pA). (b) A large area STM image of $C_8$-BTBT IL ($V_{sample}$ = -1.29 V and I = 9.08 pA). (c) A higher magnification STM image of the $C_8$-BTBT interface layer on graphene/Cu foil. ($V_{sample}$ = -0.91 V and I = 12.1 pA). The angle between the carbon chains and the benzothiophene is measured to be 140°. The arrangement of molecule is shown in inset of (c) and (d). The angle between the carbon chains and the benzothiophene is 134° calculated by DFT. The angle between benzothiophene plane and the graphene substrate is ~10°. (e) The section lines along the green lines in (c) and (d). (f) The section lines along the lines in (c) and (d). We can see a π/2 phase ascending from both lines a to c and lines a' to c'. (g) A typical *dI/dV-V* curve recorded at the IL in Fig. 2(c).

agrees with the fact that the HOMO-LUMO gap of a $C_8$-BTBT molecule is about 3.84 eV [33].

To further explore the effect of substrate flatness on the growth of high-quality single crystal molecular layers, we intentionally carry out STM measurements of the IL grown on substrate with large out-of-plane roughness (there are many wrinkles of graphene [23] and large steps of Cu foil). Figure 3(a) shows a typical STM image of our controlled experiment. Obviously, the large roughness of the substrate induces disturbance on the growth of high-quality $C_8$-BTBT IL. Small-area domains of the $C_8$-BTBT IL with different orientations are observed due to the high $C_8$-BTBT nucleation density induced by the large roughness. The relative angles between these domains usually are not multiples of 60°, which indicates that the $C_8$-BTBT molecules do not prone to growing along the zigzag or armchair directions of graphene [34-39]. According to our DFT calculation, the maximum difference of binding energy for $C_8$-BTBT along different directions of graphene is only ~ 6.8 meV/molecule [19]. The result in Fig. 3(a-c) indicates that the orientation of the $C_8$-BTBT domain is mainly determined by its nucleation center (for example, the direction of steps in the substrate). Our experimental result also points out that the height of the out-of-plane roughness is a critical parameter that determines the growth of the $C_8$-BTBT IL, as shown in Fig. 3(d). Only the steps with height larger than 0.6 nm could act as the $C_8$-BTBT nucleation center. For the substrates of graphene/$SiO_2$ and BN/$SiO_2$, the $SiO_2$ could also induce out-of-plane roughnesses with height larger than 0.6 nm [40] (also see

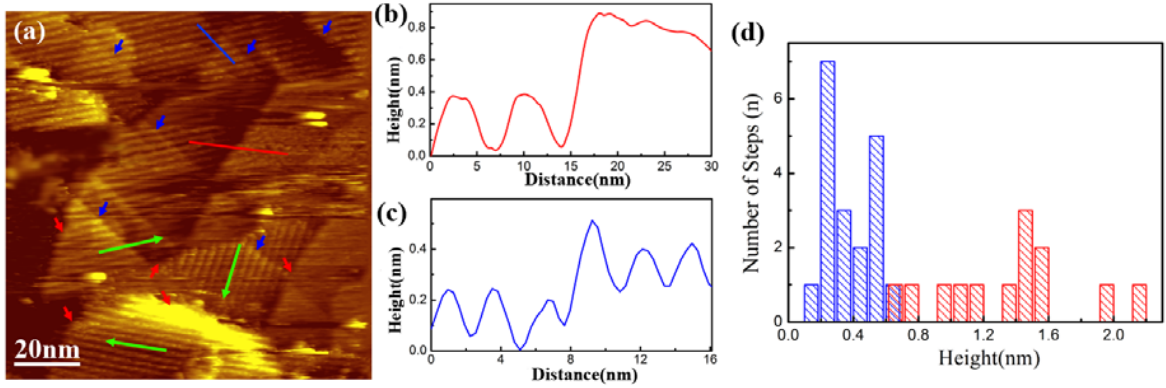

**Figure 3** (color online). (a) A typical STM image showing multiple domains of the IL ($V_{sample}$ = 1.04 V and I = 10.5 pA). The green arrows denote the orientations of three different domains. The relative angles between these domains are 21.5°, 98.0°, and 60.5°. The blue and red arrows point to out-of-plane roughness of the substrate. The growth of the IL is disturbed by the out-of-plane roughness pointed out by the red arrows but not by these indexed by the blue arrows. Two typical section lines at the positions of blue and red lines are given in (b) and (c). (d) The number of steps that do (not) affect the growth of the IL [red bars (blue bars)] versus their height measured in our experiment.

Figure S5 in Supporting Material [24] for a typical profile line on graphene/$SiO_2$), which are the nucleation centers for the growth of the $C_8$-BTBT layers. Of course, high-density of such roughnesses could induce disturbance on the growth of $C_8$-BTBT IL and is harmful to grow large-area single crystal $C_8$-BTBT IL. In our transport measurement [19], we carefully select the substrate and reduce the $C_8$-BTBT nucleation centers. With successful synthesis of large-area single crystal $C_8$-BTBT layers, we achieve a record-high charge carrier mobility ~ 10 $cm^2$/Vs in the $C_8$-BTBT 1L. This information would be important in guiding the growth of high-quality epitaxy molecular layers.

## IV. CONCLUSIONS

In conclusion, the structures of the 2D $C_8$-BTBT crystals were carefully studied using AFM and cryogenic STM. We demonstrate that the $C_8$-BTBT molecules prefer to grow layer-by-layer on graphene and BN and only out-of-plane roughness greater than 0.6 nm of the substrates could act as the $C_8$-BTBT nucleation center. The ability to grow high-quality single crystal molecular layers reported in this paper opens the way to the applications of OFETs and is expected to play an important part in future electronics.

**Acknowledgements**

We are grateful to National Key Basic Research Program of China (Grant No. 2014CB920903, No. 2013CBA01603, No. 2013CBA01604), National Science Foundation

(Grant No. 11422430, No. 11374035, No. 11474022, No. 11274154, No. 51172029, No. 61325020, No. 61261160499), the program for New Century Excellent Talents in University of the Ministry of Education of China (Grant No. NCET-13-0054), Beijing Higher Education Young Elite Teacher Project (Grant No. YETP0238), and the Fundamental Research Funds for the Central Universities.

† These authors contribute equally to this work.

* Correspondence to Lin He (helin@bnu.edu.cn) and Xinran Wang (xrwang@nju.edu.cn).